# Heat shield for de Sitter space

P. C. W. Davies, Logan Thomas, and George Zahariade

*Department of Physics and Beyond: Center for Fundamental Concepts in Science,
Arizona State University, Tempe, Arizona 85287, USA*

We investigate a quantum vacuum state for $(1+1)$-dimensional de Sitter space, corresponding to a region within a perfectly reflecting symmetric box of fixed physical size. We find that a particle detector inside the box registers zero response, implying that the walls of the box screen out the thermal effects of the de Sitter horizon. The box thus creates a "quiescent oasis" with a temperature below the de Sitter horizon temperature, an unexpected feature that opens the way to an analysis of heat and entropy exchange between the box and the horizon in the context of the generalized second law of thermodynamics. We also calculate the stress-energy-momentum tensor of the region within the box, showing that the total energy in the box is less than the same volume of de Sitter space in the presence of thermal fluctuations.

## I. INTRODUCTION

A landmark advance in theoretical physics was Hawking's discovery that black holes possess a well-defined temperature and entropy, thus establishing a link among gravitation, quantum mechanics, and thermodynamics [1,2]. Soon afterwards, it was found that black hole radiance has close analogs for accelerating observers (Rindler space) and for de Sitter space [3–5]. There are two rather different ways to explore the thermodynamic properties of these systems. One is to determine the response of a particle detector, and the other is to compute the renormalized stress-energy-momentum tensor. Early results established that a static Unruh-DeWitt detector [4,6] located far from a black hole registers a thermal response at the Hawking temperature and, consistent with this, the stress tensor corresponds to a flux of thermal radiation at that same temperature [7,8].

The situation with the Rindler and de Sitter systems is, however, less obvious. In the latter case, an inertial particle detector registers a temperature $T_{\rm dS} = H/2\pi$ where $H = \sqrt{\Lambda/3}$ is the de Sitter parameter and $\Lambda$ is the cosmological constant [5].[1] However, the stress tensor does not correspond to thermal radiation. For example, a conformally invariant scalar field in a de Sitter–invariant vacuum state (the Bunch-Davies vacuum) yields the result [9]

$$\langle T_{\mu\nu}\rangle = \frac{H^4}{960\pi^2}g_{\mu\nu} \propto g_{\mu\nu}, \qquad (1)$$

which amounts to a renormalization of $\Lambda$. By contrast, one would expect a bath of thermal radiation to be described by a stress tensor $\langle T_{\mu\nu}\rangle \propto {\rm diag}(1, 1/3, 1/3, 1/3)$. In fact, the stress tensor in a thermal vacuum at temperature $T_{\rm dS}$ actually yields

$$\langle T_{\mu\nu}\rangle_{T_{\rm dS}} = \frac{H^4}{960\pi^2}g_{\mu\nu} + \frac{H^4(H\eta)^2}{720\pi^2}{\rm diag}\left(1,\frac{1}{3},\frac{1}{3},\frac{1}{3}\right). \qquad (2)$$

Regardless of this observation, the thermodynamic properties of the quantum vacua of de Sitter space have been at the forefront of recent advances in cosmology. Indeed, they play a crucial role in the inflationary universe scenario [10–13] by providing a quantum origin to the primordial fluctuations that seed the density perturbations in the postinflationary epoch [14–16]. They are also central to extensions of the second law of thermodynamics. Bekenstein and Hawking [17–20] generalized the second law to include the transfer of energy and entropy between a black hole and its environment. This was subsequently further generalized to incorporate de Sitter horizons [5]. In fact, many detailed analyses have demonstrated that these results are robust under a wide range of circumstances (see, for example, [21] and the references contained therein). This lends weight to the idea that the thermodynamic properties of de Sitter space bear close resemblance to those of black hole spacetimes.

To explore this relationship further, an interesting thought experiment is to consider the case of a Schwarzschild black hole confined to a perfectly reflecting box. Inside the box, the black hole reaches thermodynamic equilibrium and is described by the Hartle-Hawking quantum vacuum [22]. Outside the box one may choose the Boulware vacuum state in which a static particle detector registers a zero temperature [23]. This effectively screens the exterior from the thermal effects of the black hole. In this paper we investigate the analogous situation for de Sitter space. We ask whether it is possible to screen out the thermal effects of the de Sitter horizon in the same way that a box around a black hole can screen its thermal effects from the surrounding universe. We are therefore interested to know whether one may create a "quiescent oasis" within de Sitter space with a temperature less than $T_{dS}$. To investigate this, we compute both the response of a particle detector and the stress tensor within this "oasis" region. Quantum field theory effects in de Sitter space in the presence of a spherical reflecting boundary in arbitrary dimensions have been studied by Milton and Saharian [24]. However, in their setup, the boundary is comoving and passes over the horizon at late times. Therefore de Sitter thermal effects cannot be screened. We focus instead on the case of a fixed (non-comoving) boundary and restrict the analysis to a massless scalar field in two spacetime dimensions. This enables us to exploit the conformal symmetry of the system in order to satisfy the reflecting boundary conditions. Our principal result is that both the detector response and the vacuum stress tensor within the cavity fail to register the thermal effects of de Sitter space outside the cavity. Whether all vestige of de Sitter geometry are erased at the quantum level by the chosen boundary conditions on the shell is left as an open question. Being a two-dimensional analysis, the "shell" reduces to two reflecting points symmetrically placed on either side of the origin. This is analogous to the Casimir effect, but with suitably chosen "plate" trajectories. We are confident that a four-dimensional analysis would yield similar results.

---

[1] We work in units where $\hbar = c = k_B = 1$.

The structure of the paper is as follows. In Sec. II, we set up our notations and describe the system of interest. In Sec. III, we briefly discuss the case of a reflecting box of constant comoving size, which was more extensively treated in [24]. Section IV is devoted to our main results: the computation of the response of a detector in a reflecting box of fixed "physical" size, and the study of the renormalized stress tensor inside the box. We end with a short discussion in Sec. V.

## II. SETUP AND NOTATIONS

We consider a massless scalar field $\varphi(t, x)$ propagating in the spatially flat patch of two-dimensional de Sitter space with metric

$$ds^2 = dt^2 - e^{2Ht}dx^2, \qquad (3)$$

where $H$ is the Hubble constant. (Notice that technically there is no cosmological constant $\Lambda$ in two dimensions but $H = \sqrt{\Lambda/3}$ in four dimensions.) Using the conformal time $\eta$ defined by $d\eta = -\frac{1}{H}e^{-Ht}dt$, the metric becomes

$$ds^2 = \frac{1}{(H\eta)^2}(d\eta^2 - dx^2) \qquad (4)$$

with $-\infty < t < \infty$, $-\infty < \eta < 0$. It will also be useful to introduce the coordinates $\rho, \sigma$ defined by

$$\eta - x = -\frac{1}{H}e^{-H(\rho-\sigma)}, \qquad (5)$$

$$\eta + x = -\frac{1}{H}e^{-H(\rho+\sigma)}, \qquad (6)$$

for which the metric becomes

$$ds^2 = \frac{1}{\cosh^2(H\sigma)}(d\rho^2 - d\sigma^2). \qquad (7)$$

Notice that this static coordinate system only covers the chronological past of the point $(\eta = 0, x = 0)$, which is often called the static patch of de Sitter space [see, for example, Fig. 20(e) of [9]]. In fact, $\rho$ is simply the time coordinate of the usual static coordinate system of de Sitter space, while $\sigma$ is the associated tortoise coordinate.

In the following sections we will be considering two different scenarios. We will begin by reviewing the case where $\varphi$ is confined to a region of de Sitter space between two comoving, perfectly reflecting plates.[2] Then we will move on to the case where the confining region has a fixed physical size.

## III. COMOVING MIRRORS

Consider a cavity composed of two comoving mirrors situated at $x = \pm L/2$. $L$ is thus the comoving size of the cavity. In $(\eta, x)$ coordinates the field $\varphi$ obeys the wave equation $\Box\varphi = 0$ with boundary conditions $\varphi(\eta, \pm L/2) = 0$. The normalized mode functions are

$$u_n(\eta, x) = \begin{cases} \frac{1}{\sqrt{n\pi}} e^{-\frac{in\pi\eta}{L}} \sin\left(\frac{n\pi x}{L}\right) & \text{for } n \text{ even,} \\ \frac{1}{\sqrt{n\pi}} e^{-\frac{in\pi\eta}{L}} \cos\left(\frac{n\pi x}{L}\right) & \text{for } n \text{ odd.} \end{cases} \qquad (8)$$

Here the $\eta$ dependence was chosen to match the Bunch-Davies vacuum. We are interested in computing the response of a particle detector in the cavity, so we start by writing the Wightman function as a mode sum:

---

[2]This constraint will be imposed via Dirichlet boundary conditions on the plates but we expect that the use of Neumann boundary conditions [25] would not alter the main conclusion of this work.

$$G^+(\eta, x; \eta', x') = \frac{1}{\pi} \sum_{n=0}^{\infty} \frac{1}{2n+1} e^{-\frac{i(2n+1)\pi}{L}(\eta-\eta'-i\epsilon)} \cos\left(\frac{(2n+1)\pi}{L}x\right) \cos\left(\frac{(2n+1)\pi}{L}x'\right)$$

$$+ \frac{1}{\pi} \sum_{n=1}^{\infty} \frac{1}{2n} e^{-\frac{i2n\pi}{L}(\eta-\eta'-i\epsilon)} \sin\left(\frac{2n\pi}{L}x\right) \sin\left(\frac{2n\pi}{L}x'\right). \quad (9)$$

(We will suppress the $i\epsilon$'s from now on for readability.) Carrying out the summations, we obtain

$$G^+(\eta, x; \eta', x') = \frac{1}{8\pi} [2\,\mathrm{argtanh}(e^{-\frac{i\pi}{L}(\eta-\eta'+x-x')}) + 2\,\mathrm{argtanh}(e^{-\frac{i\pi}{L}(\eta-\eta'+x+x')})$$
$$+ 2\,\mathrm{argtanh}(e^{-\frac{i\pi}{L}(\eta-\eta'-x+x')}) + 2\,\mathrm{argtanh}(e^{-\frac{i\pi}{L}(\eta-\eta'-x-x')})$$
$$+ \ln\left(1 - e^{-\frac{2i\pi}{L}(\eta-\eta'+x+x')}\right) + \ln\left(1 - e^{-\frac{2i\pi}{L}(\eta-\eta'-x-x')}\right)$$
$$- \ln\left(1 - e^{-\frac{2i\pi}{L}(\eta-\eta'+x-x')}\right) - \ln\left(1 - e^{-\frac{2i\pi}{L}(\eta-\eta'-x+x')}\right)]. \quad (10)$$

The response of a particle detector situated at the origin is obtained by first setting $x = x' = 0$ in the previous expression. This yields

$$G^+(\eta, 0; \eta', 0) = \frac{1}{\pi} \mathrm{argtanh}(e^{\frac{-i\pi}{L}(\eta-\eta')}). \quad (11)$$

In terms of the proper time $t$ of the detector this would correspond to a detector response function that is nonstationary since $\eta - \eta' = (e^{-Ht'} - e^{-Ht})/H$ is not a function of only $t - t'$. In physical terms, the detector registers a time-varying flux of particles.

Let us now turn to the calculation of the renormalized stress tensor, $\langle T_{\mu\nu} \rangle$. One may directly calculate this quantity from the Wightman function, but here we will find it more useful to quote the general result [see, e.g., Eq. (6.136) of [9]],

$$\langle T_\mu{}^\nu(u,v) \rangle = (-g)^{-1/2} \langle T_\mu{}^\nu(u,v) \rangle_M + \theta_\mu{}^\nu - \frac{1}{48\pi} R \delta_\mu{}^\nu. \quad (12)$$

Here $\langle T_\mu{}^\nu(u,v) \rangle_M$ is the renormalized stress tensor in Minkowski space, and $u = \eta - x$ and $v = \eta + x$ are the usual light-cone coordinates. $\theta$ is defined by

$$\begin{cases} \theta_{ii} = \frac{1}{24\pi} \left[ \frac{\partial_i^2 C}{C} - \frac{3}{2} \left( \frac{\partial_i C}{C} \right)^2 \right], & \text{for } i = u \text{ or } i = v, \\ \theta_{uv} = \theta_{vu} = 0, \end{cases} \quad (13)$$

where $C = [H(u+v)/2]^{-2}$ is the conformal factor of the metric in light-cone coordinates. For the scenario studied in this section $\theta_{\mu\nu} = 0$. Moreover, $\langle T_\mu{}^\nu(u,v) \rangle_M$ is simply the familiar Casimir effect result [26] and the Ricci scalar is $2H^2$. In terms of $(\eta, x)$ coordinates, we find

$$\langle T_{\mu\nu}(\eta, x) \rangle = -\frac{\pi}{24 L^2} \delta_{\mu\nu} - \frac{H^2}{24\pi} g_{\mu\nu}. \quad (14)$$

This result is readily understood: the first term is a consequence of the presence of the expanding cavity (with the energy density falling to zero at late times) while the second term is a renormalization of the cosmological constant. Notice finally that this can be obtained as the $d \to 2$ limit of the general result in [24].

## IV. FIXED MIRRORS

Consider now a cavity composed of two fixed mirrors situated at $x(t) = \pm e^{-Ht} L/2$. $L$ is thus the physical size of the cavity. In $(\eta, x)$ coordinates the motion of the mirrors is uniform with (coordinate) velocity $\mp HL/2$, i.e., $x(\eta) = \mp \eta H L/2$. Transforming to the $(\rho, \sigma)$ coordinates in (5) and (6) further simplifies the problem since the mirrors are then situated at

$$\sigma = \pm \frac{1}{2H} \ln\left(\frac{1 + HL/2}{1 - HL/2}\right) \equiv \pm \ell/2. \quad (15)$$

Notice that $L < 2/H$ to ensure that all points within the cavity are in causal contact with the origin (where our detector will be placed). In $(\rho, \sigma)$ coordinates the field $\varphi$ still obeys the wave equation $\Box \varphi = 0$. (This is due to the conformal symmetry of the wave equation in two dimensions.) However, the boundary conditions now involve $\ell$ instead of $L$: $\varphi(\rho, \pm \ell/2) = 0$. Therefore the mode functions are the same as in the previous case with $L$ replaced by $\ell$ and the whole analysis carries over unchanged. The particular (static) quantum vacuum that is picked via this procedure is analogous to the Boulware vacuum of black hole physics (see [27] for a nice discussion of different de Sitter vacua).

Generically the response function per unit time of an Unruh-DeWitt particle detector with worldline given by $x(\tau)$ (where $\tau$ is the proper time) is

$$\mathcal{F}(E) = \int_{-\infty}^{\infty} d(\Delta\tau) e^{-iE(\tau-\tau')} G^+(\tau, x(\tau), \tau', x(\tau')). \quad (16)$$

This quantifies the probability per unit time for the detector to transition from its ground state to an excited state of energy E. The Wightman function at the origin, in terms of cosmological time, is given by

$$G^+(t,0;t',0) = \frac{1}{\pi}\operatorname{argtanh}(e^{-\frac{i\pi}{\ell}(t'-t-i\epsilon)}). \quad (17)$$

This coincides with the Minkowski space Wightman function of a scalar field in a box of size $\ell$. Hence the response function is identical to that of a Minkowski space detector situated in the middle of a static cavity of size $\ell$, and this is, of course, 0. We conclude that the effect of the cavity with a reflecting boundary is to screen the thermal effects of de Sitter space.

If the detector is placed at a constant spatial coordinate, $\sigma = \bar{\sigma}$, then it is uniformly accelerating, and one might expect to find an associated Unruh temperature related to this acceleration. However, this is not so, as may be seen from the Wightman function

$$G^+(\rho,\bar{\sigma},\rho',\bar{\sigma}) = \frac{1}{4\pi}[\ln(1+e^{-\frac{i\pi}{\ell}(\rho-\rho'+2\bar{\sigma})})$$
$$+ \ln(1+e^{-\frac{i\pi}{\ell}(\rho-\rho'-2\bar{\sigma})})$$
$$- 2\ln(1-e^{-\frac{i\pi}{\ell}(\rho-\rho')})]. \quad (18)$$

In our coordinate system, the proper time of the detector is simply related to the coordinate time by $\Delta\tau = \Delta\rho/\cosh(H\bar{\sigma}) \equiv \Delta\rho/a(\bar{\sigma})$. We can rewrite the logarithms using their Taylor expansions (with the understanding that unwritten $i\epsilon$ factors ensure their convergence for all $\Delta\tau$). The response function integral then becomes trivial:

$$\mathcal{F}(E) = \frac{1}{4\pi}\int_{-\infty}^{\infty}d(\Delta\tau)e^{-iE\Delta\tau}$$
$$\times \sum_{n=1}^{\infty}\frac{1}{n}\left[2\cos\left(\frac{2n\pi}{\ell}\bar{\sigma}\right)e^{-\frac{i\pi}{\ell}a(\bar{\sigma})\Delta\tau} - (-1)^n e^{-\frac{i\pi}{\ell}a(\bar{\sigma})\Delta\tau}\right]$$
$$= \frac{1}{2\pi}\sum_{n=1}^{\infty}\frac{1}{n}\delta\left(E+\frac{n\pi a(\bar{\sigma})}{\ell}\right)\left[\cos\left(\frac{2\pi n}{\ell}\bar{\sigma}\right)-(-1)^n\right],$$
$$(19)$$

which vanishes for all $E > 0$. Thus, even an accelerating detector registers zero temperature.

We can now use Eq. (12) to compute the stress tensor. Using the light-cone coordinates on the static patch $\bar{u} = \rho - \sigma$, $\bar{v} = \rho + \sigma$, we see that $C = \cosh^{-2}(H(\bar{v}-\bar{u})/2)$, and thus

$$\theta_{\bar{u}\bar{u}} = \theta_{\bar{v}\bar{v}} = -\frac{H^2}{48\pi}. \quad (20)$$

Transforming this into $(\rho,\sigma)$ coordinates gives

$$\langle T_{\mu\nu}(\rho,\sigma)\rangle = -\frac{1}{24}\left[\frac{H^2}{2\pi}+\frac{\pi}{\ell^2}\right]\delta_{\mu\nu} - \frac{H^2}{24\pi}g_{\mu\nu}. \quad (21)$$

Note that raising one index will introduce a factor of $1/C = \cosh^2(H\sigma)$, which diverges on the horizon ($\sigma \to \infty$). The first term on the right of Eq. (21) will thus diverge as the edges of the box approach the horizon. This effect is characteristic of the static vacuum of de Sitter space; it is analogous to the case of the Schwarzschild black hole where the energy momentum tensor is also known to diverge on the horizon in the Boulware vacuum.

We may now integrate $\langle T^{00}\rangle$ with respect to the invariant measure, over a constant $\rho$ slice to find the total energy in the cavity:

$$E = \int_{-\ell/2}^{\ell/2}\frac{d\sigma}{\cosh(H\sigma)}\left(-\frac{1}{24}\cosh^4(H\sigma)\right.$$
$$\times\left[\frac{H^2}{2\pi}+\frac{\pi}{\ell^2}\right] - \frac{H^2}{24\pi}\cosh^2(H\sigma)\right)$$
$$= -\left[\frac{H^2}{2\pi}+\frac{\pi}{\ell^2}\right]\frac{5+\cosh(H\ell)}{72H}\sinh\left(\frac{H\ell}{2}\right)$$
$$- \frac{H}{12\pi}\sinh\left(\frac{H\ell}{2}\right). \quad (22)$$

This result has two components. The first term includes the effect of the mode discretization due to the finite cavity. The second term is the energy that would be present in that region of space even in the absence of a box, i.e., if the box had transparent walls and the quantum state was the Bunch-Davies vacuum. It is thus clear that the energy of the vacuum state investigated here is less than that of the Bunch-Davies vacuum, consistent with the Minkowski space limit of negative Casimir energy.

We now turn to examine the thermal state in this cavity. The result in Eq. (19) can be written as a relationship between the detector response in the cavity and the detector response of a stationary detector in a box of fixed size in Minkowski space,

$$\mathcal{F}_{dS}(E) = \frac{1}{\cosh(H\bar{\sigma})}\mathcal{F}_M(E/\cosh(H\bar{\sigma})), \quad (23)$$

where $\mathcal{F}_{dS}$ is the detector response in our cavity and $\mathcal{F}_M$ is the detector response in the analogous box in Minkowski. Such a relationship is only possible because the mode functions are of the same functional form, and the proper time along our chosen trajectory is a constant multiple of the coordinate time. These conditions are state independent; i.e., Green's function in some state in Minkowski space has the same functional form as that of the analogous state in dS. Therefore, at finite temperature, the detector response is identical to the detector response in the thermal state of a conducting box in flat spacetime, up to the factors of $\cosh(H\bar{\sigma})$ appearing in Eq. (23).

For a massless scalar in two dimensions, the stress tensor is

$$T_{\mu\nu} = \partial_\mu \phi \partial_\nu \phi - \frac{1}{2} g_{\mu\nu} g^{\alpha\beta} \partial_\alpha \phi \partial_\beta \phi. \tag{24}$$

In order to compute the stress tensor in the thermal state, we need to compute $\langle \partial_\mu \phi \partial_\nu \phi \rangle$,

$$\langle \partial_\mu \phi \partial_\nu \phi \rangle_\beta = \lim_{\rho' \to \rho} \lim_{\sigma' \to \sigma} \frac{1}{2}(\partial_\mu \partial_{\nu'} + \partial_{\mu'} \partial_\nu) \frac{1}{2} G_\beta^{(1)}(\rho, \sigma; \rho', \sigma'). \tag{25}$$

The thermal anticommutator propagator, $G_\beta^{(1)}$, can be computed as [9]

$$G_\beta^{(1)}(\rho, \sigma; \rho', \sigma') = \sum_{m=-\infty}^{\infty} [G^+(\eta + im\beta, \sigma; \eta', \sigma') + G^{+*}(\eta + im\beta, \sigma; \eta', \sigma')]. \tag{26}$$

We will focus on the diagonal components of $\langle \partial_\mu \phi \partial_\nu \phi \rangle_\beta$ separately. The off-diagonal components are 0,

$$\langle (\partial_\rho \phi)^2 \rangle_\beta = \frac{\pi \cos^2(\frac{\pi\sigma}{\ell})}{16\ell^2} \sum_{m=-\infty}^{\infty} \frac{-1 + 2\cos(\frac{2\pi\sigma}{\ell}) + 2\cosh(\frac{m\pi\beta}{\ell}) + \cosh(\frac{2m\pi\beta}{\ell})}{\sinh^2(\frac{m\pi\beta}{2\ell})[\cos(\frac{2\pi\sigma}{\ell}) + \cosh(\frac{m\pi\beta}{\ell})]^2}, \tag{27}$$

$$\langle (\partial_\sigma \phi)^2 \rangle_\beta = -\langle (\partial_\rho \phi)^2 \rangle_\beta + \frac{\pi}{16\ell^2} \sum_{m=-\infty}^{\infty} \frac{1}{\sinh^2(\frac{m\pi\beta}{2\ell})}. \tag{28}$$

Although there is a singular term in each sum when $m = 0$, this corresponds to the 0-temperature contribution which we have already calculated. If we combine Eqs. (27) and (28) in the fashion of Eq. (24), the complicated sum disappears,

$$\langle T_{\mu\nu} \rangle_\beta = \langle T_{\mu\nu} \rangle_\infty + \left[\frac{\pi}{8\ell^2} \sum_{m=1}^{\infty} \frac{1}{\sinh^2(\frac{m\pi\beta}{2\ell})}\right] \delta_{\mu\nu}, \tag{29}$$

where we have separated out the zero-temperature $m = 0$ term and used the even symmetry of the summand to write the sum over the positive integers. Notably, the thermal stress tensor is time independent.

## V. CONCLUSION

In this paper, we investigated the properties of the vacuum state associated with a scalar field confined within a region of fixed size (box) in de Sitter space, with Dirichlet boundary conditions. A particle detector fails to detect any particles, registering 0 temperature. The thermal properties of this state, as opposed to that of the surrounding de Sitter spacetime, show that the effect of the cavity is to create an oasis that screens out the thermal properties of the horizon. This immediately begs the question of whether *any* of the properties of de Sitter space survive in the cavity; in particular, do particles still "feel" the de Sitter expansion of the underlying spacetime? The answer is subtle. Indeed, if particles are treated as classical test masses, then they will, of course, be insensitive to the peculiar quantum vacuum studied here and will travel along geodesics of the classical de Sitter geometry, globally moving away from each other. However, even in this case, a gas of noninteracting classical particles, or billiard balls, would likely not feel the presence of the expanding background on average since the gravitational redshift would be compensated by the Doppler effect due to reflections at the boundaries of the fixed-physical-size box. The situation is on the contrary more nuanced when considering that the particles are, in fact, quantum (as they should be), i.e., localized excitations of the field or, in other words, wave packets viewed as superpositions of field modes. In our case the modes should be chosen to correspond to the particular vacuum that we are considering, which would render the wave packets insensitive to the classical geometry. (But, of course, we are faced here with the usual ambiguities related to the definition of particles in curved spacetime.)

These subtleties notwithstanding, we expect our main result to qualitatively hold in higher dimensional de Sitter space for a conformally coupled scalar field. In more than two dimensions, the static patch of de Sitter is no longer conformally related to Minkowski space, but instead to Rindler space. The calculations presented here would be related to calculations of the stress tensor and detector response of an accelerating observer in a discretized version of the Rindler vacuum. A Rindler observer does not detect excitations of the Rindler vacuum [9], and we expect the same to be true within a reflecting shell. Moreover, the presence of additional time-independent conformal factors cannot prevent the detector response function from vanishing. However, we also calculated the stress-energy tensor within the fixed-physical-size box and showed that its total energy is less than what it would be in the absence of any reflecting boundaries. This is not expected to hold in higher dimensions; in the case where the radius of the cavity is much smaller than the de Sitter radius, we expect the energy to be larger [28].

We also expect the results to hold for conformally invariant higher-spin fermionic or bosonic fields, albeit at the expense of additional computational complexity. However, it is still unclear whether the analog of the Boulware vacuum described in this work exists for the case of nonzero mass or more generally when conformal symmetry is lost. These results seem to indicate that massless conformally coupled quantum fields cannot be used to determine the nonzero spacetime curvature of de Sitter space. However, this appears to be a feature of the conformal symmetry of the model which implies that massive quantum fields could be used to probe the geometry.

The thermal nature of de Sitter space, although analogous to that of black holes, is nevertheless somewhat problematic. Although de Sitter horizons have a well-defined temperature, they do not radiate energy; de Sitter space does not "evaporate" as do black holes. The de Sitter horizon entropy appears no different from black hole horizon entropy [21], and the two may be traded, leading to a generalized second law of thermodynamics, but the interpretation of horizon entropy as "information lost" makes little sense in de Sitter space. An open question is whether the "heat" of de Sitter space can be "mined" by a demonlike mechanism such as a Szilárd engine. To operate, such a mechanism must be isolated from the de Sitter horizon temperature. In this paper, we have demonstrated that such a zero-temperature enclave may be established. An interesting avenue of future investigation would therefore be the study of heat exchange processes between the zero temperature interior of the box and the de Sitter horizon.

## ACKNOWLEDGMENTS

We thank S. M. Carroll, S. A. Fulling, A. Ottewill, M. Parikh, and M. E. Tegmark for helpful comments. During this work L. T. and G. Z. were supported by the Foundational Questions Institute (FQXi). L. T. is also supported by the U.S. Department of Energy, under Grant No. DE-SC0019470. G. Z. also acknowledges financial support from *Moogsoft* and the Archelon Time Science Fellowship.